\documentclass[12pt,preprint]{aastex}

\usepackage{epsfig,graphicx,natbib}
\usepackage{amssymb}
\usepackage{amsfonts}
\usepackage{amsmath}
\usepackage{color}
\usepackage{lscape}

\shorttitle{Low-Mass Star Formation Timescales} \shortauthors{Banerji et al.}

\begin{document}


\title{Timescales for Low-Mass Star Formation in Extragalactic Environments: Implications for the Stellar IMF}


\author{M. Banerji\altaffilmark{1},
  S. Viti\altaffilmark{1},
  D.A. Williams\altaffilmark{1}
  and J.M.C. Rawlings\altaffilmark{1}}

\email{mbanerji@star.ucl.ac.uk}


\altaffiltext{1}{Department of Physics and Astronomy, University
  College London, Gower Street, London WC1E 6BT, UK.}


\begin{abstract}

We investigate the physical and chemical conditions necessary for low-mass star formation in extragalactic environments by calculating various characteristic timescales associated with star formation for a range of initial conditions. The balance of these timescales indicates whether low-mass star formation is enhanced or inhibited under certain physical conditions. In this study, we consider timescales for free-fall, cooling, freeze-out, desorption, chemistry and ambipolar diffusion and their variations with changes in the gas density, metallicity, cosmic ray ionisation rate and FUV radiation field strength. We find that extragalactic systems with high FUV radiation field strengths and high cosmic ray fluxes considered at a range of metallicities, are likely to have enhanced low-mass star formation unless the magnetic pressure is sufficient to halt collapse. Our results indicate that this is only likely to be the case for high-redshift galaxies approaching solar metallicities. Unless this is true for all high-redshift sources, this study finds little evidence for a high-mass biased IMF at high redshifts.
\end{abstract}





\section{Introduction}

Recent multi-wavelength studies of galaxies have permitted estimates to be made of the evolution of the space density of the massive star formation rate \citep{Hopkins:SFR,Lilly:SFR,Madau:SFR}. While these rates are at present poorly constrained, there is currently no information on the formation at high redshift of low mass stars, a process that in the Milky Way is known to accompany the formation of massive stars. The low mass stars may be optically insiginificant, but they modify the physics of the host galaxies rather significantly. In particular, the shape of the stellar initial mass function (IMF) and its slope as specified by the ratio of low-mass to high-mass stars, are crucial for determining the star formation history and stellar mass density of a galaxy. There has been much interest in varying the stellar IMF \citep{Elmegreen:IMF,Wilkins:IMF,Dave:08,VanDokkum:08} and recent studies have suggested that the IMF must be top-heavy at high redshifts to explain the counts of faint submillimeter galaxies and the luminosity function of Lyman break galaxies \citep{Baugh05:IMF}. However, this has by no means been proven true for all high redshift environments. 

Studies of low mass star formation in the Milky Way have shown that it arises in the interplay of a number of competing processes including gravitational collapse and thermal and magnetic support. Thermal support at the low temperatures of interstellar clouds is determined by the cooling mechanisms, while magnetic support is controlled by the ionization fraction; both are largely determined by chemistry which itself is moderated by cosmic ray and photon fluxes, and metallicity. A consideration of the timescales appropriate for the Milky Way indicates - as it should - that low mass star formation takes place under Milky Way conditions. However, the physical properties in high redshift galaxies may depart significantly from the Milky Way values. In this paper, therefore, we compute in a rather crude way how these relevant timescales may vary when these physical conditions are modified from Milky Way values. We use the results to determine the regions of parameter space in which low mass star formation is unlikely to occur. By doing so, we are able to investigate claims of a high-mass biased IMF at high redshifts.   

In $\S$ \ref{sec:timescales} we list the relevant physical processes and give estimates of the timescales. The cooling functions and ionization fractions are determined from a complex model of interstellar chemistry described in $\S$ \ref{sec:model}. Our results are described in $\S$ \ref{sec:results}, and some general conclusions regarding low mass star formation at high redshifts are made in $\S$ \ref{sec:conclusions}.

\section{Timescales for Star Formation}

\label{sec:timescales}

Star formation is a complex process. While gravity is the main driver, there are many other mechanisms that compete to aid and oppose the collapse of molecular clouds. Cooling by molecular radiation leads to thermally induced collapse of a molecular cloud and if the timescale associated with this process is longer than the free-fall time, the collapse will be halted. The rate of cooling in turn is determined by the abundances of the various different molecular coolants such as CO and so we can see the role that chemistry plays. The chemistry itself is driven by cosmic ray ionisation of molecular hydrogen. In some circumstances, the surfaces of dust grains can become coated with ice due to the \textit{freeze out} of different species onto the grains. This process is effective in removing coolant molecules from the gas phase. Energy released from $H_2$ formation can liberate the \textit{frozen} species into their gaseous phase. Interstellar magnetic fields too can affect the collapse of molecular clouds. The ions are coupled to field lines and neutral atoms to the ions via collisions. The fractional ionisation therefore determines the magnetic pressure and the level of ion-neutral friction which in turn can halt collapse.  

In the local universe, the timescales associated with these various competing processes are found to be roughly of the same order of magnitude - $\sim 10^{5}$ yrs \citep{Williams:astrochem_intro} for a core density of $\sim10^4$cm$^{-3}$. This balance of timescales is clearly conducive to star formation. By exploring the change in these timescales for a range of initial conditions, we can effectively explore different regions of redshift space and hope to learn something about when and how the first low mass stars started to form. 

\subsection{Collapse}

\label{sec:collapse}

Low mass star formation essentially occurs by the collapse of a molecular cloud under its own weight accompanied by the dissipation of gravitational potential energy. In any situation where the star formation is not externally triggered by a violent event, for example the passage of a shock, the actual collapse time will be somewhat longer than the free-fall time due to turbulence within the cloud. In this work, we use the free-fall time as a useful under-estimate of the typical collapse time for a molecular cloud undergoing quiescent star formation.   

We can relate the free-fall time of a collapsing core to the number density of hydrogen nuclei as follows:

\begin{equation}
t_{ff}=\sqrt{\frac{3\pi}{16G\rho}}=0.75\times10^8/{(n_H)}^{\frac{1}{2}} yr
\label{eq:tff}
\end{equation}

where $n_H=(n(H)+2n(H_{2}))$ is the local number density of hydrogen nuclei in cm$^{-3}$ \citep{DysonWilliams:ISM}.

\subsection{Cooling}

\label{sec:cooling}

In order for a cloud of gas to continue to collapse and form a star, the cloud must continue to lose energy, otherwise, the high temperatures and densities within the cloud would halt collapse before the formation of a star could occur. Energy loss occurs via different atomic and molecular transitions, while at high densities, collisions between hot gas and cooler dust grains can also be efficient in cooling the gas. The rate at which energy is lost depends on the cooling function, $\Lambda_{tot}/$ erg cm $^{-3}$ s$^{-1}$, which in turn depends on the abundance of various different molecular coolants. The total cooling time is then given by:

\begin{equation}
t_{cool}=\frac{\frac{5}{2}n_HkT}{\xi\Lambda_{tot}} yr
\label{eq:tcool}
\end{equation}

\noindent where $k$ is the Boltzmann constant, $T$ is the gas temperature and $\xi$ is the metallicity in units of solar metallicity. The quantity, $\frac{5}{2}n_HkT$ represents the total energy, both thermal and internal, of the system and $\xi\Lambda_{tot}$ is the total interstellar cooling rate. 

Note that our chemical models described in $\S$ \ref{sec:model}, do not include the effects of removing molecular coolants from the gas phase by freeze-out. If freeze-out occurs, the cooling timescale will be underestimated in regions of parameter space where there is significant ice-formation. 

\subsection{Freeze-out}

\label{sec:freezeout} 

A detailed calculation of the rate per unit volume at which a species freezes out can be found in \citet{Rawlings:freezeout}. This quantity depends on the relative abundance of dust grains compared to hydrogen nuclei, the physical properties of the grain as well as the species whose freeze-out is being considered, and electrostatic effects to take into account the fact that ionic species will freeze-out at a different rate to neutral species. Due to the presence of more than one grain population in most systems, the grain properties need to be averaged over the grain size distribution. \citet{Rawlings:freezeout} relate this size distribution to the depletion coefficient, $D$ which depends on the abundance of very small grains and on the metallicity.     

Taking these assumptions into consideration, the freeze-out timescale can be derived from the rate of freeze-out given by \citet{Rawlings:freezeout}:

\begin{equation}
t_{fo}=\frac{0.9\times10^6{(m_{X}/28)}^\frac{1}{2}}{(T/10)^\frac{1}{2}(n_H/10^{4})D} yr
\label{eq:tfo}
\end{equation}

where $m_{X}$ is the molecular mass of the species whose freeze-out is being considered and $T$ is the gas temperature.

\subsection{Desorption}

\label{sec:desorption}

Desorption is the process by which molecules in ices on the surface of dust grains, are returned to the ISM. The mechanism through which heat is generated for this desorption, is still debated \citep{Roberts:desorption}. In this work, we consider only one mechanism for desorption, namely exothermic reactions occurring on the grain surface. More specifically, the liberation of energy from $H_2$ formation on dust grains results in a proportion of the frozen molecules on the dust grain being returned to the gas phase. 

We calculate the desorption rate of CO which is the most abundant molecular species besides hydrogen. This rate is:

\begin{equation}
\frac{d n(CO)}{dt}\vert_{des}=3\times10^{-17}nn_{H}\xi\gamma\sqrt{\frac{T}{100}}
\label{eq:dtdes}
\end{equation}

The rate depends on the metallicity which scales linearly with the dust to gas ratio, the abundance of both molecular and atomic hydrogen, the temperature, $T$ and the number of CO molecules desorbed for every $H_2$ formed, $\gamma$. The coefficient $3\times10^{-17}$ represents the canonical $H_2$ formation rate in the Milky Way.  

From this expression, we can calculate a typical timescale associated with desorption: 

\begin{equation}
t_{des}=\frac{1\times10^9}{[n_Hn(H)(T/100)^\frac{1}{2}\gamma\xi]} yr
\label{eq:tdes}
\end{equation}

where $n_H$ is the local number density of hydrogen nuclei and $n(H)$ is the column density of hydrogen atoms only. The number of CO molecules desorbed per $H_2$ formation, $\gamma$ can be constrained by considering a pre-stellar core in the Milky Way like L1689B where significant depletion of CO molecules has been directly observed \citep{Redman:L1689B}. Assuming for such an object with $n_H\sim10^4-10^5$cm$^{-3}$, that the timescale for freeze-out is less than half the timescale for desorption gives an upper limit on $\gamma$ of $\sim$0.25. We use $\gamma=0.2$ throughout this work and assume that this number does not change in high redshift systems. This is plausible as we expect the microscopic properties of the dust in high redshift galaxies to be fairly similar to those in the Milky Way. Both the freeze-out and desorption timescales depend on these microscopic properties.  

\subsection{Ambipolar Diffusion}

\label{sec:ambipolar}

Magnetic fields may play an important role in supporting dense molecular clouds and many of the fragments within them. The magnetic field lines are coupled directly to the ions and indirectly to the neutral matter through frequent ion-neutral collisions. At low ionisations, the frequency of the ion-neutral collisions becomes less regular and the field lines decouple from the neutral matter thereby allowing the matter to drift in response to the gravitational potential. This process is known as ambipolar diffusion and plays an important role in determining the rate of quiescent star formation. 

The characteristic timescale associated with this process of decoupling the field from the neutral matter will depend on the relative drift velocities of the ions and neutral particles. \citet{Hartquist:89} define the ambipolar diffusion timescale as follows:

\begin{equation}
t_{amb}=4\times 10^5(x_i/10^{-8}) yr
\label{eq:tamb}
\end{equation}

This is a function only of the fractional ionisation, $x_i$ within the cloud and is consistent with the more detailed treatment of \citet{Mouschovias:79}. However, in the earlier paper, the ambipolar diffusion timescale is shown to be a function of position within the cloud. Therefore variations in the degree of ionisation as a function of cloud depth become important. For the purposes of our rather crude calculations, we neglect this positional dependence. Eq. \ref{eq:tamb} therefore corresponds to the \citet{Mouschovias:79} timescale at an intermediate position towards the interior of the core. It should be noted that this timescale is only relevant for magnetically sub-critical cores.  

\subsection{Ion-Molecule Chemistry}

\label{sec:chemistry}

The chemical timescale is the time taken to create molecular coolants from the initially atomic gas assuming that nearly all of the hydrogen is molecular. In other words, it is the time taken to ionise sufficient hydrogen to allow complete chemical conversion of C and O to molecular form. Cosmic rays are responsible for ionising the hydrogen. The relative abundances of carbon and oxygen will scale with the metallicity, $\xi$. The chemical timescale is

\begin{equation}
t_{chem}\simeq \frac{3\xi(n_{C}^{tot}+n_{O}^{tot})}{n(H_2)\zeta}= \frac{5\times10^6\xi}{\zeta/1\times10^{-17}} yr
\label{eq:tim}
\end{equation}

The chemistry is important for providing the coolant molecules and therefore must not be too long if star formation is to occur. The analytical expression for the chemical timescale represented by Eq. \ref{eq:tim} depends only on the metallicity and cosmic ray ionisation rate and is rather crude. The chemical timescale is less relevant in cores than in the parent cloud which may be collapsing from a diffuse to a dark state as there isn't much change in the chemistry once cores begin to form.

\section{Modelling}

\label{sec:model}

In order to model the various different timescales discussed in $\S$ \ref{sec:timescales} we use a time and depth-dependent chemical code for molecular clouds and photon-dominated regions to solve simultaneously for the chemistry, thermal balance and radiative transfer within a cloud. The code models the gas phase chemistry inside the cloud which is treated as a 1D semi-infinite slab of gas illuminated by a uni-directional flux of FUV photons incident on its outer surface \citep{Bell:Xfactor1}. Differences in model geometry are unlikely to affect the chemistry significantly enough to influence our rather qualitative results (See \citet{Rollig:PDR} for a PDR benchmarking code comparison). The FUV radiation field is taken to be the standard interstellar radiation field \citep{Draine:ISRF} multiplied by a factor of $\chi$ where $\chi=1.7$ corresponds to a flux of $1.6\times10^{-3}$ erg s$^{-1}$ cm$^{-2}$, appropriate for the unshielded interstellar medium in the Milky Way. The radiation field strength declines exponentially with increasing depth within the cloud and therefore depends on the visual extinction $A_v$. No grain chemistry is included and hence the freeze-out and desorption of species is ignored in our models. Various different heating and cooling mechanisms are invoked within the code. Photoejection of electrons from dust grains and polycyclic aromatic hydrocarbons (PAHs) and FUV pumping and photo-dissociation of $H_2$ are the dominant heating mechanisms while the gas is mainly cooled through emission from collisionally excited atoms and molecules and by interactions with the cooler dust grains. The gas temperature is determined at each depth point by using an iterative procedure to balance the total heating and cooling rates. Although the chemical model is also time dependent, we consider all outputs at steady state, assumed to correspond to a time of 1Gyr.

The code takes as its inputs the number density of hydrogen nuclei, $n_H$, the metallicity, $\xi$, the cosmic ray ionisation rate, $\zeta$, the strength of the incident FUV radiation field, $\chi$ and the turbulent velocity, $v$ within the cloud. As we are considering quiescent low-mass star formation only, we fix the turbulent velocity at a value of $v=1.5$ km s$^{-1}$. We will consider variations in the other parameters and implications for the star formation timescales in $\S$ \ref{sec:results}.  

The chemistry consists of a chemical network containing 83 species. These species interact in over 1300 reactions, whose rates are taken from the UMIST99 database \citep{LeTeuff:UMIST99}. Initially, hydrogen is assumed to be in atomic form. 

Given these inputs, the model calculates chemical abundances for the different species, emission line strengths and gas temperatures as a function of the visual extinction, $A_v$ for every time step.  

\section{Results}

\label{sec:results}

In this section we describe how the various timescales associated with low mass star formation described in detail in $\S$ \ref{sec:timescales} vary as we change the free parameters. We consider variations in the number density of hydrogen nuclei, $n_H$, the metallicity, $\xi$, the cosmic ray ionisation rate, $\zeta$ and the strength of the incident FUV radiation field, $\chi$.  When varying any of the four parameters, the others are fixed at their \textit{local} values namely $n_H=10^5$ cm$^{-3}$, $\xi=\xi_\odot$, $\zeta=10^{-17}$ s$^{-1}$ and $\chi=1.7$. All outputs are considered at steady state corresponding to a time of 1Gyr and at two depths corresponding to $A_v\simeq3$ and $A_v\simeq10$. Outputs of the code include the cooling function, $\Lambda_{tot}$, the fractional ion abundance, $x_i$, the gas temperature $T$ and the column density of hydrogen atoms $n(H)$ and these values can be substituted into equations \ref{eq:tcool}, \ref{eq:tfo}, \ref{eq:tdes} and \ref{eq:tamb}  to calculate the cooling, freeze-out, desorption and ambipolar diffusion timescales. 

For each set of initial conditions, we make some references to extragalactic objects that may exhibit these properties and hence are able to infer if low mass star formation is likely to be encouraged or inhibited in such objects. We emphasise that our calculations are crude in nature and instructive only in terms of observing general trends.

\subsection{Varying the Gas Density}

\begin{figure}
\plotone{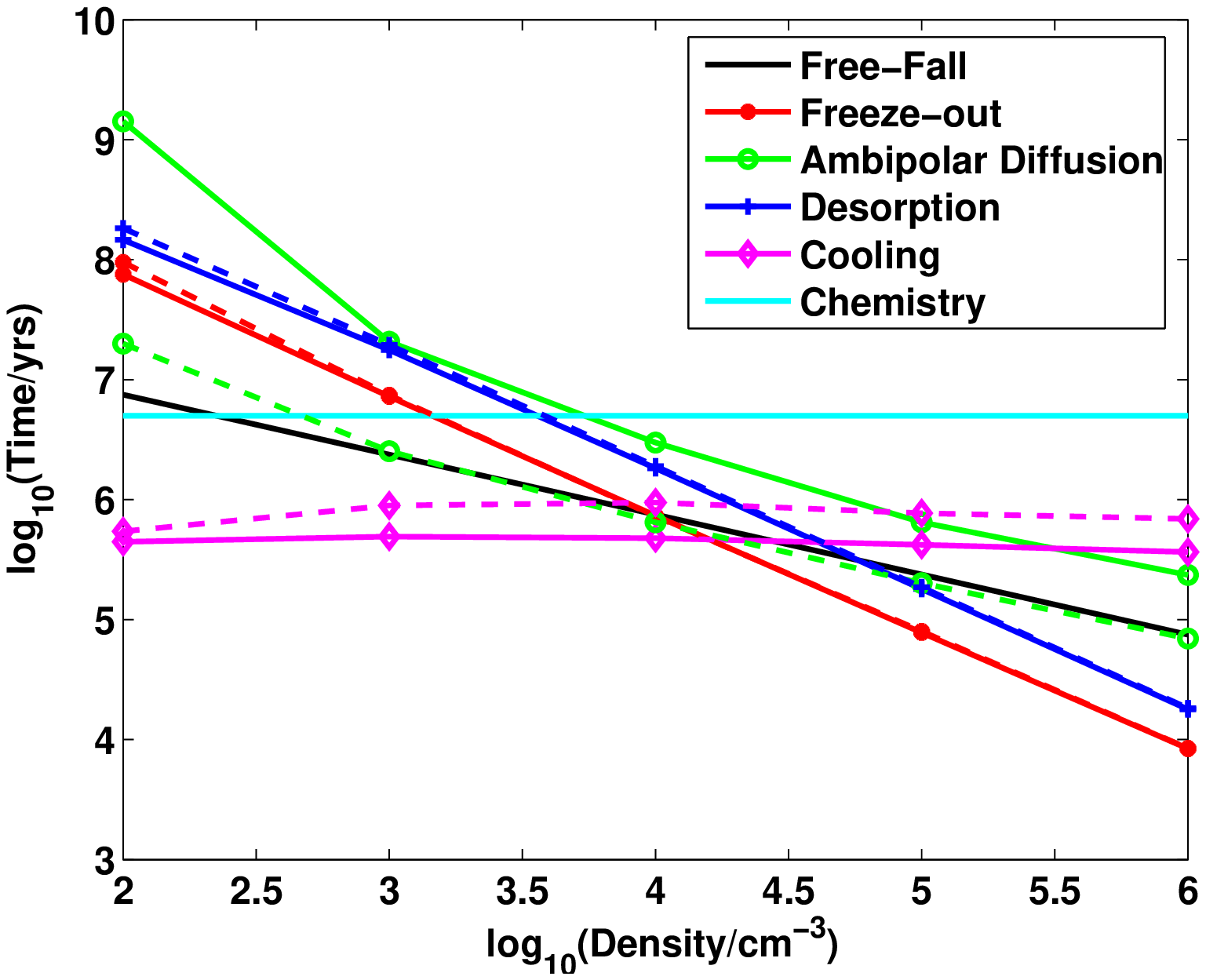}
\caption{Variation of timescales relevant to star formation with number density of hydrogen at $A_v\sim$3 (solid lines) and $A_v\sim$10 (dotted lines). The metallicity, cosmic ray ionisation rate and incident FUV radiation intensity are fixed at $\xi=\xi_\odot$, $\zeta=10^{-17}s^{-1}$ and $\chi=1.7$.}
\label{fig:n}
\end{figure}

 We vary the hydrogen nuclei number density between 100 cm$^{-3}$ and 10$^6$ cm$^{-3}$ in this study and look at the effect of this variation on the low mass star formation timescales. The results are summarised in  Figure \ref{fig:n}.

As expected, the free-fall time decreases with increasing number density converging to $\sim 10^6$yrs at 10$^4$ cm$^{-3}$. At this density, all the different timescales converge to roughly the same order of magnitude. Thus both freeze-out and desorption proceed with desorption a slightly slower process so that molecular ices do form. Desorption maintains a population of molecular coolants in the gas phase. The free-fall time is comparable to the cooling time, thermally induced collapse occurs and stars are able to form. The ambipolar diffusion timescale is comparable to the dynamical free-fall time and star-formation can occur in magnetically sub-critical couds. We note that these conditions are appropriate for a Milky Way type galaxy and therefore our approach indicates as expected that low mass star formation is reasonably efficient in Milky Way type systems.

However, at very high densities, the free-fall time is shorter than the cooling time and the cores will heat up during the collapse which is then arrested.


The freeze-out timescale and desorption timescale are comparable and both decrease with increasing density at the same rate suggesting that molecular ices are able to form at all densities between 100 cm$^{-3}$ and 10$^6$ cm$^{-3}$. This also implies that the cooling time is systematically under-estimated as ice formation, which is not included in our chemical model, removes molecular coolants from the gas phase at all densities. 

Looking deeper into the cloud, i.e. going from $A_v\simeq3$ to $A_v\simeq10$ does not seem to affect significantly the freeze-out or desorption timescales. However, the cooling time is slightly larger at higher visual extinctions as the cooling function decreases with increasing $A_v$. This is probably due to increased radiation trapping in the cooling lines. 


The ambipolar diffusion timescale is much longer than the other timescales at very low densities. This suggests as expected that star formation cannot occur in clouds with $n_H<10^3$ cm$^{-3}$ that are magnetically sub-critical as the field is very slow to decouple.

\subsection{Varying the Metallicity}

\begin{figure}
\plotone{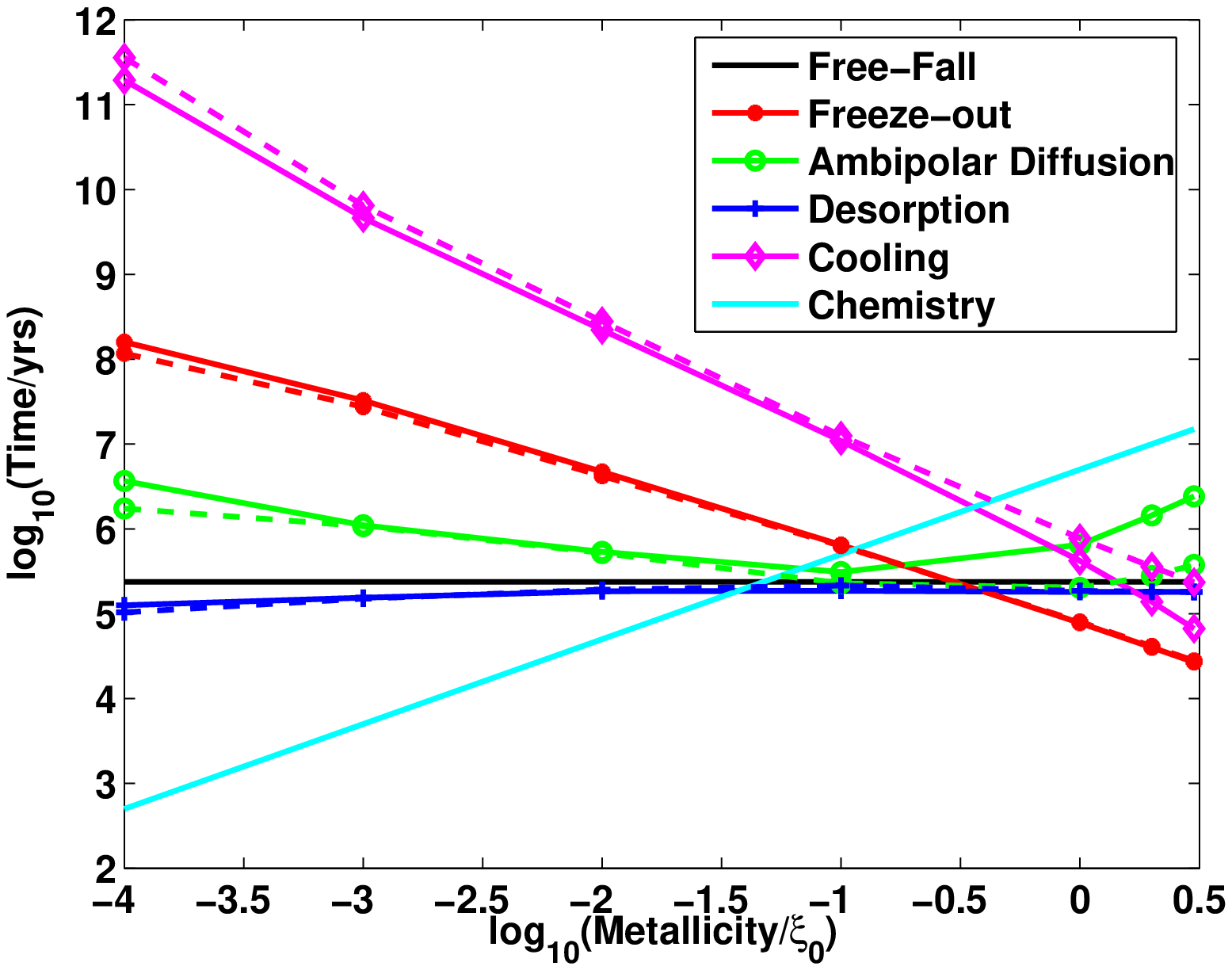}
\caption{Variation of timescales relevant to star formation with metallicity at $A_v\sim$3 (solid lines) and $A_v\sim$10 (dotted lines). The number density of hydrogen, cosmic ray ionisation rate and incident FUV radiation intensity are fixed at $n=10^5cm^{-3}$, $\zeta=10^{-17}s^{-1}$ and $\chi=1.7$.}
\label{fig:xi}
\end{figure}

The metallicity, $\xi$, is another important physical property that will affect low mass star formation. This parameter is particularly important for modelling high redshift systems as the primordial gas from which these objects formed is likely to be less chemically enriched than the interstellar medium in the Milky Way in some cases, and will therefore have lower metallicities. As we are interested in low metallicity objects that formed early in the history of the Universe, we consider metallicities as low as $10^{-4}\xi_{\odot}$. We also consider metallicities enhanced above solar values to 3$\xi_{\odot}$. In our model, we assume that the elemental abundances of all metals scale linearly with metallicity. We also assume that the dust to gas mass ratio scales linearly with metallicity. The results from this study are summarised in Figure \ref{fig:xi}.  

As expected, the cooling time falls with increasing metallicity due to the increased abundance of molecular coolants. We can see that at metallicities below $\xi=0.1\xi_\odot$,the cooling time is much longer than the free-fall time and collapse and star-formation are likely to be impeded by gas pressure. The freeze-out time also decreases with increasing metallicity due to the presence of more dust grains but the desorption time is independent of metallicity. At low metallicities, freeze-out is slower than desorption so ice-formation is inhibited. Temperatures however remain high due to the long cooling timescales. 

The ambipolar diffusion timescale remains fairly insensitive to the metallicity. The fractional ionisation is much higher at low visual extinctions so the ambipolar diffusion timescale falls with increasing depth into the cloud. Note that the visual extinction scales linearly with metallicity so high visual extinctions in low metallicity clouds will correspond to larger physical distances within the cloud. Hence, the calculations for $A_v\simeq10$ are probably unfeasible for low metallicity clouds. We have considered outputs at $A_v \simeq 7.5$ and $A_v \simeq 9.5$ for $\xi=10^{-4}\xi_\odot$ and $\xi=10^{-3}\xi_\odot$ respectively. As the cloud is likely to be optically thick even at these lower depths, the chemistry is unlikely to be very different to the $A_v\simeq10$ case.

The chemical timescale increases linearly with metallicity as expected. Low metallicity systems therefore convert a significant fraction of available elements into molecular coolants.

\subsection{Varying the Cosmic Ray Ionisation Rate}

\label{sec:cr}

Although little is known about the origin of extragalactic cosmic rays, they are very important in ionising hydrogen and driving the chemistry during star formation. The level of ionisation within a dark cloud may determine whether or not the cloud is stable against gravitational collapse if magnetic fields are present.

\begin{figure}
\plotone{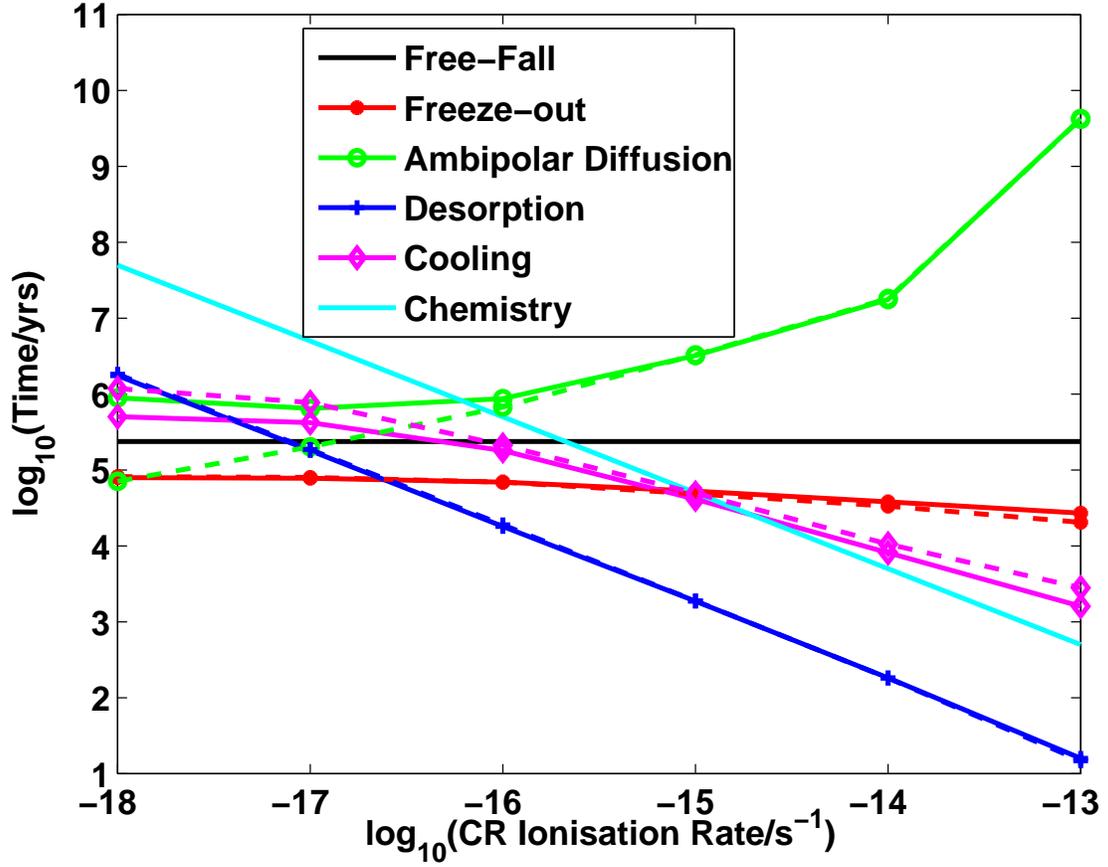}
\caption{Variation of timescales relevant to star formation with the cosmic ray ionisation rate at $A_v\sim$3 (solid lines) and $A_v\sim$10 (dotted lines). The number density of hydrogen, metallicity and incident FUV radiation intensity are fixed at $n=10^5cm^{-3}$, $\xi=\xi_{0}$ and $\chi=1.7$.}
\label{fig:zeta}
\end{figure}

An estimate for the cosmic ray ionisation rate in the Galaxy is $\sim10^{-17}$ s$^{-1}$ \citep{Hartquist:zeta}. In this study, we vary the cosmic ray ionisation rate between $10^{-18}$ s$^{-1}$ and $10^{-13}$ s$^{-1}$ in order to mimic both quiescent star forming systems such as those in our own galaxy as well as starburst regions and AGN which are expected to have cosmic ray fluxes of the order of $10^{-15}$s$^{-1}$ to $10^{-13}$s$^{-1}$ \citep{Meijerink:cr,Suchkov:m82cr,Bradford:n253cr}. The results of varying the cosmic ray ionisation rate are summarised in Figure \ref{fig:zeta}. 

As the cosmic ray ionisation rate is increased, the cooling time falls and at $\zeta \sim10^{-16}$ s$^{-1}$, it drops below the free-fall time. Cores cool more and more quickly and are therefore able to collapse to form stars. The chemical timescale is small at high values of the cosmic ray ionisation rate explaining the low cooling times as molecular coolants are quick to form in such circumstances.  

The freeze-out time is greater than the desorption time and the propensity for ice-formation is reduced thereby allowing molecular coolants to remain in the gas phase to effectively cool the collapsing gas. Desorption is therefore unlikely to be a significant process. However, formally, the desorption time becomes shorter as expected with increasing cosmic ray ionisation rate as more and more hydrogen molecules are ionised thereby increasing the abundance of atomic hydrogen as well as the temperature of the gas. If ices could be formed, they would therefore be quick to evaporate off the dust grains once again replenishing the gas phase with molecular coolants and encouraging star formation.

The ambipolar diffusion timescale becomes very large as the fractional ionisation increases and the magnetic field is slow to be released from the collapsing core. Based on the results of \citet{Crutcher:mag} and \citet{Athreya:B}, the magnetic field strength, B of molecular clouds varies between 1$\mu$Gauss and 100$\mu$Gauss for densities between 1cm$^{-3}$ and 10$^5$ cm$^{-3}$. This field strength, B scales with the density as $B\propto (n_H)^\kappa$ where $\frac{1}{3}\leq\kappa\leq\frac{1}{2}$ \citep{Mouschovias:kappa} is consistent with observations. In this study, based on various observations \citep{Crutcher:mag, Bourke:mag,Beck:B,Kim:B} we consider $\kappa \sim 0.4$. We can calculate the ratio of the magnetic pressure and thermal pressure, $\alpha$ using Eq. \ref{eq:magpres}:

\begin{equation}
\alpha=\frac{B^2}{8\pi n_H k T}
\label{eq:magpres}
\end{equation}

\noindent For $\alpha>>1$ the magnetic pressure dominates and for $\alpha<<1$, the thermal pressure dominates. For high cosmic ray fluxes and at a constant density of $n_H$=10$^5$ cm$^{-3}$, we compute $\alpha \sim 0.4$. This suggests that the magnetic pressure may in some cases be sufficient to halt the collapse but in other cases, depending on the temperature of the core and the exact value of $\kappa$, low-mass star formation can proceed unimpeded.

\subsection{Varying the FUV Radiation Field}

\begin{figure}
\plotone{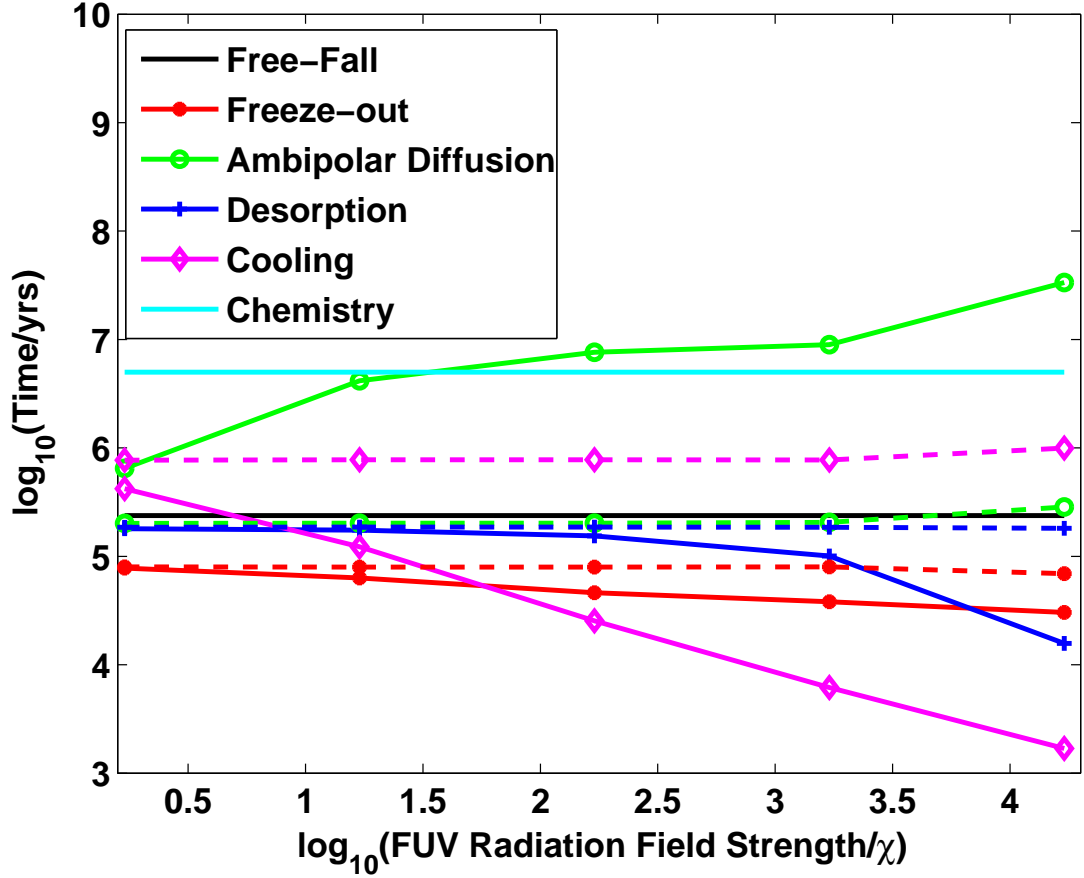}
\caption{Variation of timescales relevant to star formation with the FUV radiation field strength at $A_v\sim$3 (solid lines) and $A_v\sim$10 (dotted lines). The number density of hydrogen, metallicity and cosmic ray ionisation rate are fixed at $n=10^5cm^{-3}$, $\xi=\xi_{0}$ and $\zeta=1\times10^{-17} s^{-1}$.}
\label{fig:G}
\end{figure}

The FUV radiation field is a significant source of heating of neutral interstellar gas. As explained in $\S$ \ref{sec:model}, the radiation field is taken to be the standard interstellar radiation field \citep{Draine:ISRF} multiplied by a factor of $\chi$ where $\chi=1.7$ corresponds to a flux of $1.6\times10^{-3}$ erg s$^{-1}$ cm$^{-2}$. The FUV radiation field strength is linked to the number of massive stars in the Galaxy and given the presence of many massive stars at high redshift \citep{Abel:Firststars,Bromm:Firststars} is likely to be large in these systems. In this study, we vary $\chi$ between 1.7 and 1.7$\times 10^4$ and study the effect on our star formation timescales. The results are summarised in Figure \ref{fig:G}. 

At $A_v \simeq 10$, the FUV radiation is sufficiently attenuated to not affect the chemistry within the cloud and all our timescales therefore remain fairly constant with FUV flux strength. At lower visual extinctions, we find that the cooling time is almost always smaller than the free-fall time and collapse and star formation are therefore encouraged. The freeze-out time is also almost always smaller than the desorption timescale except at very high FUV radiation field strengths. This suggests that the formation of icy mantles on dust grains is possible even at FUV radiation field strengths of $\chi=1.7\times10^3$ at $A_v\sim3$. The desorption time and freeze-out time are fairly comparable at all radiation field strengths so any icy mantles deposited on the dust grain are quick to evaporate. At very high FUV flux strengths, the desorption time is shorter than the freeze-out time and ice formation is inhibited. Both these processes mean that the ISM is always rich in molecular coolants at high FUV radiation field strengths and collapse is encouraged.

The ambipolar diffusion timescale is very large for high radiation field strengths and at low $A_v$. This once again suggests that magnetically super-critical cores will be stable against collapse. However, at high $A_v$ where the radiation field cannot penetrate the cloud, the ambipolar diffusion timescale is comparable to the dynamical timescale and the magnetical criticality is less significant. 

As we are linking both the FUV radiation field strengths and the cosmic ray ionisation rate with massive stars in this rather simplistic analysis, it is not surprising that both parameters have similar effects on the star formation timescales. We conclude that the presence of massive stars resulting in an elevated FUV flux strength, serves to encourage quiescent low-mass star formation in every possible way at low $A_v$ provided the residual magnetic pressure implied by the long ambipolar diffusion timescale, does not halt collapse. We compute the ratio of the magnetic to thermal pressure for $\chi=1.7\times10^4$ to be $\alpha \sim 0.5$ and conclude that the magnetic pressure may in some cases be sufficient to halt collapse.

\begin{table*}
  \caption{Star Formation Timescales at High-Redshift} \label{tab:highz}
  \begin{center}
    \begin{tabular}{c| cc | cc | cc }
      \textbf{Timescale} & \multicolumn{2}{c|}{\textbf{Model I}} & \multicolumn{2}{|c|}{\textbf{Model II}} & \multicolumn{2}{|c}{\textbf{Model III}} \\
      \hline
      Years & $A_v\sim3$ & $A_v\sim10$ & $A_v\sim3$ & $A_v\sim10$ & $A_v\sim3$ & $A_v\sim10$ \\
      \hline \hline
      Free-Fall & $7.50\times10^5$ & $7.50\times10^5$ & $2.37\times10^5$ & $2.37\times10^5$ &$2.37\times10^5$ & $2.37\times10^5$\\
      Cooling & $6.76\times10^5$ & $7.98\times10^5$ & $4.31\times10^5$ & $6.75\times10^6$ & $3.29\times10^3$ & $1.07\times10^4$ \\
      Freeze-out & $1.68\times10^6$ & $1.35\times10^6$ & $5.22\times10^5$ & $1.38\times10^5$ & $4.01\times10^4$ & $3.93\times10^4$ \\
      Desorption & $3.11\times10^2$ & $1.94\times10^2$ & $1.11\times10^2$ & $2.35\times10^1$ & $1.44\times10^2$ & $1.81\times10^2$ \\
      Ambipolar Diffusion & $7.14\times10^{10}$ & $8.58\times10^{10}$ & $1.76\times10^8$ & $2.45\times10^7$ & $1.43\times10^9$ & $1.94\times10^7$ \\
      Chemistry & $2.50\times10^1$ & $2.50\times10^1$ & $2.50\times10^2$ & $2.50\times10^2$ & $5.00\times10^3$ & $5.00\times10^3$\\
      \hline
    \end{tabular}    \vspace{2mm}
  \end{center}
\end{table*}

\subsection{Modelling a High Redshift Object}

\citet{Bayet:2008} have identified the Cloverleaf ($z\simeq2.6$) and APM08279 ($z\simeq3.9$) as possible models for high redshift sources. Although the physical properties of these sources are not well constrained, molecular detections suggest that they are actively forming stars \citep{Cloverleaf:SF,Gao:04,Wagg:APM}. We expect such objects to have a range of different metallicities with galaxies with significant observed dust obscuration such as HCM6A ($z=6.56$,\citet{Chary:05}) likely to be more chemically enriched. The presence of many massive stars in these systems will also result in high FUV radiation field strengths and cosmic ray ionisation rates. Collapse is likely to occur at moderately high densities due to the lack of available molecular coolants compared to a Milky Way type object.

In order to model such a system, we calculate the star formation timescales for three different models - (I) $n=10^4$ cm$^{-3}$, $\xi=0.05\xi_{\odot}$, $\zeta=10^{-13}$ s$^{-1}$ and $\chi=1.7\times10^4$ , (II) $n=10^5$ cm$^{-3}$, $\xi=0.05\xi_{\odot}$, $\zeta=10^{-14}$ s$^{-1}$ and $\chi=1.7\times10^3$ and (III) $n=10^5$ cm$^{-3}$, $\xi=\xi_{\odot}$, $\zeta=10^{-14}$ s$^{-1}$ and $\chi=1.7\times10^3$ - at depths corresponding to $A_v\sim3$ and $A_v\sim10$. As previously stated, the presence of many massive stars in high redshift systems justifies our choice of elevated FUV radiation field strengths when modelling such systems. In order to mimic starburst galaxies with high cosmic ray fluxes such as M82 \citep{Suchkov:m82cr} and NGC253 \citep{Bradford:n253cr} at high redshift, we adopt high cosmic ray ionisation rates. \citet{Meijerink:cr} estimate a cosmic ray ionization rate of $5\times10^{-15}$s$^{-1}$ for a star formation rate of $\sim$100M$_\odot$ yr$^{-1}$ and we expect high redshift galaxies with higher star formation rates to have elevated cosmic ray fluxes. Finally, there is evidence that high redshift galaxies exist at a range of metallicities with galaxies with significant observed dust obscuration likely to be more chemically enriched and we therefore consider both solar and sub-solar metallicities. Our results are summarised in Table \ref{tab:highz}.

We can see that for the first model, the free-fall time is comparable to the cooling time and cores can collapse to form stars even at depths as low as $A_v\sim3$. The formation of icy mantles is likely to be inhibited in such systems as the freeze-out time is very large compared to the desorption time.

The second model has slightly lower cosmic ray fluxes and FUV radiation field strengths but is considered to be at a higher density. In this case, the free-fall timescale is shorter than the cooling time at low $A_v$ and cores can collapse to form stars but at $A_v\sim10$, the cooling time is larger and the propensity for low mass star formation is reduced. Once again, the formation of icy mantles is inhibited in such a system.

The third model also has elevated FUV radiation field strengths and cosmic ray fluxes but is now considered at solar metallicity. The cooling time is now significantly lower than the free-fall time at all depths and low-mass star formation is unimpeded by thermal pressure while ice formation still remains inhibited. 

In all three cases, the ambipolar diffusion timescale is very large so the residual field may be able to halt collapse. Using the temperatures and densities of these model runs, we calculate the thermal and magnetic pressure for each of the models in the same way as in $\S$ \ref{sec:cr} and their ratio, $\alpha$. In both the sub-solar metallicity models, we compute $\alpha\sim0.05$ so the thermal pressure dominates over the magnetic pressure. The cooling times are small compared to the dynamical free-fall time, particularly at low $A_v$ and low-mass star formation is encouraged. The third model at solar metallicity has $\alpha\sim0.6$ suggesting that in this case the magnetic pressure may be sufficient to halt collapse. We conclude that recent evidence for a top-heavy IMF is only supported by this study if high-redshift galaxies are considered to be at solar metallicities in which case the magnetic pressure may halt the collapse and formation of low-mass stars.

\section{Conclusions}

\label{sec:conclusions}

In this work we have investigated the variation of certain characteristic timescales associated with low mass star formation with physical properties of the collapsing cloud. By looking at the balance of these timescales we are able to determine the conditions necessary for the formation of low mass stars and comment on the form of the stellar initial mass function in extragalactic environments. Despite the crudity of our calculations, we observe the following trends:

(i) Increasing the density results in the cooling time becoming longer than the free-fall time, so that cores tend to heat up and star formation is inhibited. Assuming that the microscopic properties of dust are similar in extragalactic environments to that in our own galaxy, we find that the formation of molecular ices is likely to occur at all densities between 100 cm$^{-3}$ and 10$^6$ cm$^{-3}$. The ambipolar diffusion timescale is comparable to the free-fall time except at very low densities where it is much longer, suggesting that low mass star-formation is allowed in magnetically sub-critical clouds with densities above $10^3$ cm$^{-3}$. 

(ii) Low metallicity systems are characterised by long cooling times compared to the free-fall timescale. Collapsing cores tend to heat up thus inhibiting star formation and collapse. Due to the low dust grain number density, the desorption time is much shorter than the freeze-out time in low metallicity systems and there is less ice formation than in the Milky Way.

(iii) High cosmic ray ionisation rates and high FUV radiation field strengths are associated with the presence of massive stars. Such systems show rapid cooling times compared to their free-fall timescales and star formation is unimpeded by thermal pressure. The propensity for ice formation is reduced in systems with cosmic ray ionisation rates that are much larger than those in the Milky Way. Freeze-out occurs over long timescales and if any ice was formed, it would evaporate quickly; molecular coolants would be returned to the ISM further increasing the star formation efficiency. The freeze-out and desorption timescales are comparable for systems at almost all FUV radiation field strengths so that icy mantles are able to form except at very high values of $\chi$. The ambipolar diffusion timescale is very long for high values of the cosmic ray ionisation rate and radiation field strength. This implies that star-formation may not be viable in clouds that are magnetically sub-critical under these conditions.

(iv) High redshift systems are characterised by high cosmic ray fluxes and FUV radiation field strengths and can be considered to exist at a range of metallicities. All such systems will show enhanced low-mass star formation with the propensity for star formation increasing with the metallicity of the galaxy provided thermal pressure dominates the magnetic pressure. This is certainly shown to be the case in active high redshift galaxies at sub-solar metallicities. However, in active galaxies at solar metallicities, the magnetic field can impede the formation of low-mass stars resulting in a high-mass biased stellar IMF. 

While this study is relatively crude in nature, it is instructive in terms of making some predictions about essentially unobservable low-mass star formation in extragalactic environments that are different from our own Galaxy. We have noted the kinds of systems in which star formation is likely to be enhanced or inhibited by considering changes in various timescales relevant for star formation. We found among other things that low-mass star formation is likely to be encouraged in host galaxies of massive stars and AGN and inhibited in low metallicity systems with low cosmic ray fluxes and FUV radiation field strengths. Our study finds tenuous evidence for a top-heavy stellar IMF at high redshifts but only if galaxies at high redshift are considered to be at near solar metallicities. In the future, we hope to extend this study by varying other free parameters such as the individual elemental abundances. We also hope to be able to relate this study to empirical estimations of star formation rates in high redshift systems.

\begin{acknowledgements}
We acknowledge the referee for insightful comments and helping to place the paper within a broader context. MB is supported by an STFC studentship. 
\end{acknowledgements}

\newpage

\newpage

\bibliography{}

\begin{thebibliography}{36}
\expandafter\ifx\csname natexlab\endcsname\relax\def\natexlab#1{#1}\fi

\bibitem[{{Abel} {et~al.}(2002){Abel}, {Bryan}, \& {Norman}}]{Abel:Firststars}
{Abel}, T., {Bryan}, G.~L., \& {Norman}, M.~L. 2002, Science, 295, 93

\bibitem[{{Athreya} {et~al.}(1998){Athreya}, {Kapahi}, {McCarthy}, \& {van
  Breugel}}]{Athreya:B}
{Athreya}, R.~M., {Kapahi}, V.~K., {McCarthy}, P.~J., \& {van Breugel}, W.
  1998, \aap, 329, 809

\bibitem[{{Baugh} {et~al.}(2005){Baugh}, {Lacey}, {Frenk}, {Granato}, {Silva},
  {Bressan}, {Benson}, \& {Cole}}]{Baugh05:IMF}
{Baugh}, C.~M., {Lacey}, C.~G., {Frenk}, C.~S., {Granato}, G.~L., {Silva}, L.,
  {Bressan}, A., {Benson}, A.~J., \& {Cole}, S. 2005, \mnras, 356, 1191

\bibitem[{{Bayet} {et~al.}(2008){Bayet}, {Viti}, {Williams}, \&
  {Rawlings}}]{Bayet:2008}
{Bayet}, E., {Viti}, S., {Williams}, D.~A., \& {Rawlings}, J.~M.~C. 2008, \apj,
  676, 978

\bibitem[{{Beck} {et~al.}(2005){Beck}, {Fletcher}, {Shukurov}, {Snodin},
  {Sokoloff}, {Ehle}, {Moss}, \& {Shoutenkov}}]{Beck:B}
{Beck}, R., {Fletcher}, A., {Shukurov}, A., {Snodin}, A., {Sokoloff}, D.~D.,
  {Ehle}, M., {Moss}, D., \& {Shoutenkov}, V. 2005, \aap, 444, 739

\bibitem[{{Bell} {et~al.}(2006){Bell}, {Roueff}, {Viti}, \&
  {Williams}}]{Bell:Xfactor1}
{Bell}, T.~A., {Roueff}, E., {Viti}, S., \& {Williams}, D.~A. 2006, \mnras,
  371, 1865

\bibitem[{{Bourke} {et~al.}(2001){Bourke}, {Myers}, {Robinson}, \&
  {Hyland}}]{Bourke:mag}
{Bourke}, T.~L., {Myers}, P.~C., {Robinson}, G., \& {Hyland}, A.~R. 2001, \apj,
  554, 916

\bibitem[{{Bradford} {et~al.}(2003){Bradford}, {Nikola}, {Stacey}, {Bolatto},
  {Jackson}, {Savage}, {Davidson}, \& {Higdon}}]{Bradford:n253cr}
{Bradford}, C.~M., {Nikola}, T., {Stacey}, G.~J., {Bolatto}, A.~D., {Jackson},
  J.~M., {Savage}, M.~L., {Davidson}, J.~A., \& {Higdon}, S.~J. 2003, \apj,
  586, 891

\bibitem[{{Bromm} {et~al.}(1999){Bromm}, {Coppi}, \&
  {Larson}}]{Bromm:Firststars}
{Bromm}, V., {Coppi}, P.~S., \& {Larson}, R.~B. 1999, \apjl, 527, L5

\bibitem[{{Chary} {et~al.}(2005){Chary}, {Stern}, \& {Eisenhardt}}]{Chary:05}
{Chary}, R.-R., {Stern}, D., \& {Eisenhardt}, P. 2005, \apjl, 635, L5

\bibitem[{{Crutcher} {et~al.}(1987){Crutcher}, {Troland}, \&
  {Kazes}}]{Crutcher:mag}
{Crutcher}, R.~M., {Troland}, T.~H., \& {Kazes}, I. 1987, \aap, 181, 119

\bibitem[{{Dav{\'e}}(2008)}]{Dave:08}
{Dav{\'e}}, R. 2008, \mnras, 385, 147

\bibitem[{{Draine}(1978)}]{Draine:ISRF}
{Draine}, B.~T. 1978, \apjs, 36, 595

\bibitem[{{Dyson} \& {Williams}(1997)}]{DysonWilliams:ISM}
{Dyson}, J.~E. \& {Williams}, D.~A. 1997, {The physics of the interstellar
  medium} (The physics of the interstellar medium.~ Edition: 2nd ed.~Publisher:
  Bristol: Institute of Physics Publishing, 1997.~Edited by J.~E.~Dyson and
  D.~A.~Williams.~Series: The graduate series in astronomy.~ISBN: 0750303069)

\bibitem[{{Elmegreen}(2008)}]{Elmegreen:IMF}
{Elmegreen}, B.~G. 2008, ArXiv e-prints

\bibitem[{{Gao} \& {Solomon}(2004)}]{Gao:04}
{Gao}, Y. \& {Solomon}, P.~M. 2004, \apjs, 152, 63

\bibitem[{{Hartquist} {et~al.}(1978){Hartquist}, {Doyle}, \&
  {Dalgarno}}]{Hartquist:zeta}
{Hartquist}, T.~W., {Doyle}, H.~T., \& {Dalgarno}, A. 1978, \aap, 68, 65

\bibitem[{{Hartquist} \& {Williams}(1989)}]{Hartquist:89}
{Hartquist}, T.~W. \& {Williams}, D.~A. 1989, \mnras, 241, 417

\bibitem[{{Hopkins} \& {Beacom}(2006)}]{Hopkins:SFR}
{Hopkins}, A.~M. \& {Beacom}, J.~F. 2006, \apj, 651, 142

\bibitem[{{Kim} {et~al.}(1990){Kim}, {Kronberg}, {Dewdney}, \&
  {Landecker}}]{Kim:B}
{Kim}, K.-T., {Kronberg}, P.~P., {Dewdney}, P.~E., \& {Landecker}, T.~L. 1990,
  \apj, 355, 29

\bibitem[{{Le Teuff} {et~al.}(2000){Le Teuff}, {Millar}, \&
  {Markwick}}]{LeTeuff:UMIST99}
{Le Teuff}, Y.~H., {Millar}, T.~J., \& {Markwick}, A.~J. 2000, \aaps, 146, 157

\bibitem[{{Lilly} {et~al.}(1996){Lilly}, {Le Fevre}, {Hammer}, \&
  {Crampton}}]{Lilly:SFR}
{Lilly}, S.~J., {Le Fevre}, O., {Hammer}, F., \& {Crampton}, D. 1996, \apjl,
  460, L1+

\bibitem[{{Lutz} {et~al.}(2007){Lutz}, {Sturm}, {Tacconi}, {Valiante},
  {Schweitzer}, {Netzer}, {Maiolino}, {Andreani}, {Shemmer}, \&
  {Veilleux}}]{Cloverleaf:SF}
{Lutz}, D., {Sturm}, E., {Tacconi}, L.~J., {Valiante}, E., {Schweitzer}, M.,
  {Netzer}, H., {Maiolino}, R., {Andreani}, P., {Shemmer}, O., \& {Veilleux},
  S. 2007, \apjl, 661, L25

\bibitem[{{Madau} {et~al.}(1996){Madau}, {Ferguson}, {Dickinson}, {Giavalisco},
  {Steidel}, \& {Fruchter}}]{Madau:SFR}
{Madau}, P., {Ferguson}, H.~C., {Dickinson}, M.~E., {Giavalisco}, M.,
  {Steidel}, C.~C., \& {Fruchter}, A. 1996, \mnras, 283, 1388

\bibitem[{{Meijerink} {et~al.}(2006){Meijerink}, {Spaans}, \&
  {Israel}}]{Meijerink:cr}
{Meijerink}, R., {Spaans}, M., \& {Israel}, F.~P. 2006, \apjl, 650, L103

\bibitem[{{Mouschovias}(1976)}]{Mouschovias:kappa}
{Mouschovias}, T.~C. 1976, \apj, 207, 141

\bibitem[{{Mouschovias}(1979)}]{Mouschovias:79}
---. 1979, \apj, 228, 475

\bibitem[{{Rawlings} {et~al.}(1992){Rawlings}, {Hartquist}, {Menten}, \&
  {Williams}}]{Rawlings:freezeout}
{Rawlings}, J.~M.~C., {Hartquist}, T.~W., {Menten}, K.~M., \& {Williams}, D.~A.
  1992, \mnras, 255, 471

\bibitem[{{Redman} {et~al.}(2002){Redman}, {Rawlings}, {Nutter},
  {Ward-Thompson}, \& {Williams}}]{Redman:L1689B}
{Redman}, M.~P., {Rawlings}, J.~M.~C., {Nutter}, D.~J., {Ward-Thompson}, D., \&
  {Williams}, D.~A. 2002, \mnras, 337, L17

\bibitem[{{Roberts} {et~al.}(2007){Roberts}, {Rawlings}, {Viti}, \&
  {Williams}}]{Roberts:desorption}
{Roberts}, J.~F., {Rawlings}, J.~M.~C., {Viti}, S., \& {Williams}, D.~A. 2007,
  \mnras, 382, 733

\bibitem[{{R{\"o}llig} {et~al.}(2007){R{\"o}llig}, {Abel}, {Bell}, {Bensch},
  {Black}, {Ferland}, {Jonkheid}, {Kamp}, {Kaufman}, {Le Bourlot}, {Le Petit},
  {Meijerink}, {Morata}, {Ossenkopf}, {Roueff}, {Shaw}, {Spaans}, {Sternberg},
  {Stutzki}, {Thi}, {van Dishoeck}, {van Hoof}, {Viti}, \&
  {Wolfire}}]{Rollig:PDR}
{R{\"o}llig}, M., {Abel}, N.~P., {Bell}, T., {Bensch}, F., {Black}, J.,
  {Ferland}, G.~J., {Jonkheid}, B., {Kamp}, I., {Kaufman}, M.~J., {Le Bourlot},
  J., {Le Petit}, F., {Meijerink}, R., {Morata}, O., {Ossenkopf}, V., {Roueff},
  E., {Shaw}, G., {Spaans}, M., {Sternberg}, A., {Stutzki}, J., {Thi}, W.-F.,
  {van Dishoeck}, E.~F., {van Hoof}, P.~A.~M., {Viti}, S., \& {Wolfire}, M.~G.
  2007, \aap, 467, 187

\bibitem[{{Suchkov} {et~al.}(1993){Suchkov}, {Allen}, \&
  {Heckman}}]{Suchkov:m82cr}
{Suchkov}, A., {Allen}, R.~J., \& {Heckman}, T.~M. 1993, \apj, 413, 542

\bibitem[{{van Dokkum}(2008)}]{VanDokkum:08}
{van Dokkum}, P.~G. 2008, \apj, 674, 29

\bibitem[{{Wagg} {et~al.}(2005){Wagg}, {Wilner}, {Neri}, {Downes}, \&
  {Wiklind}}]{Wagg:APM}
{Wagg}, J., {Wilner}, D.~J., {Neri}, R., {Downes}, D., \& {Wiklind}, T. 2005,
  \apjl, 634, L13

\bibitem[{{Wilkins} {et~al.}(2008){Wilkins}, {Hopkins}, {Trentham}, \&
  {Tojeiro}}]{Wilkins:IMF}
{Wilkins}, S.~M., {Hopkins}, A.~M., {Trentham}, N., \& {Tojeiro}, R. 2008,
  ArXiv e-prints

\bibitem[{{Williams}(1998)}]{Williams:astrochem_intro}
{Williams}, D.~A. 1998, in Chemistry and Physics of Molecules and Grains in
  Space. Faraday Discussions No. 109, 1--13

\end{thebibliography}

\end{document}